\documentclass[reprint,twocolumn]{revtex4-1}
\usepackage{graphicx}%
\usepackage{dcolumn}%
\usepackage{amssymb}
\usepackage{amsfonts}
\usepackage{amsmath}
\usepackage{amssymb}
\usepackage{color} 
\usepackage{mathrsfs}
\usepackage{slashed}
\usepackage[breaklinks]{hyperref}
\usepackage{tikz}
\usepackage[margin=2cm,twoside]{geometry}

\newcommand{\pa}{\partial}
\newcommand{\unit}{\mathbb{I}}

\newcommand{\nn}{\nonumber}
\renewcommand{\slash}{ \not}

 \def\be{\begin{equation}}
 \def\ee{\end{equation}}
 \def\bea{\begin{eqnarray}}
 \def\eea{\end{eqnarray}}
 \def\bean{\begin{eqnarray*}}
 \def\eean{\end{eqnarray*}}
 \def\gsim{\mathrel{\rlap{\lower0.2em\hbox{$\sim$}}\raise0.2em\hbox{$>$}}}
 \def\ksim{\mathrel{\rlap{\lower0.2em\hbox{$\sim$}}\raise0.2em\hbox{$<$}}}
 \def\kg{\mathrel{\rlap{\lower0.25em\hbox{$>$}}\raise0.25em\hbox{$<$}}}

\renewcommand{\slash}{ \not}
\newcommand{\pvec}{{\bf p}}
\newcommand{\kvec}{{\bf k}}

\newcommand{\dtilde}[1]{\frac{d^3 #1}{(2\pi)^3}}

\usetikzlibrary{decorations.pathmorphing}
\usetikzlibrary{decorations.markings}
\usetikzlibrary{decorations.pathreplacing,calc}

\begin{document}

\title{The phase diagram  of the Polyakov-Nambu-Jona-Lasinio approach for finite chemical potentials}

\author{D. Fuseau$^1$,  T. Steinert$^2$  and J. Aichelin$^{1,3}$}
\affiliation{$^1$ SUBATECH, University of Nantes, IMT Atlantique, IN2P3/CNRS
4 rue Alfred Kastler, 44307 Nantes cedex 3, France}
\affiliation{$^2$ Institut fur Theoretische Physik, Heinrich Buff Rfing 16, 35392 Giessen, Germany} 
\affiliation{$^3$ Frankfurt Institute for Advanced Studies, Ruth Moufang Str. 1, 60438 Frankfurt, Germany}
\date{\today}

\begin{abstract} \noindent
We extend the SU(3) (Polyakov) Nambu Jona-Lasinio in two ways: We introduce the next to leading order contribution (in $N_c$)  in the partition function. This contribution contains explicit mesonic terms. We introduce a coupling between the gluon field and the quark degrees of freedom which goes beyond a simple rescaling of the critical temperature.  With both these improvements we can reproduce, for vanishing chemical potentials, the lattice results for the thermal properties of a strongly interacting system like pressure, energy density, entropy density, interaction measure and the speed of sound. Also the expansion parameter towards small but finite chemical potentials agrees with the lattice results. Extending the
calculations to finite chemical potentials (what does not require any new parameter) we find a first order phase transition up to a critical end point of  $T_{CEP}= 110\  MeV$ and $\mu_q = 320\ MeV$. We calculate the mass of mesons and baryons as a function of temperature and chemical potential and the transition between the hadronic and the chirally restored phase. These calculations provide an equation of state in the whole $T,\mu$ plane an essential ingredient  for dynamical calculations of ultra-relativistic heavy ion collisions but also for the physics of neutron stars  and neutron star collisions.
\end{abstract}


\maketitle

\section{Introduction}
The study of the phase diagram of strongly interacting matter has recently gained a lot of interest. This is on the one side due to the new NICA (Dubna, Russia) and FAIR (Darmstadt, Germany) facilities which are presently under construction. They will allow high precision studies in the region of a large baryon chemical potential, $\mu$, a region of the (T,$\mu$, T being the temperature) phase diagram which is presently inaccessible. On the other side the new field of simulations of  collisions of neutron stars, triggered by the observation of gravitational waves from such events, needs as input this phase diagram for a large range of densities and temperatures \cite{Bose:2017jvk}.

At zero baryochemical potential the thermodynamical quantities of strongly interacting matter have been calculated in lattice gauge calculations which gained over the years in precision and where the results of different groups agree now within the error bars \cite{Borsanyi:2013bia,Bazavov:2014pvz}. These results have been compared with experimental results at RHIC (Brookhaven National Laboratory) and LHC (CERN) in a number of ways. This includes the observation that  the multiplicity of the observed hadrons can be well described using a statistical model with a very small baryochemical potential and a temperature close to the  critical temperature of the lattice calculation, defined as the inflection point of the pressure as a function of the temperature \cite{Braun-Munzinger:2015hba,Stachel:2013zma,Cleymans:2005xv}.  Hydrodynamical calculations or kinetic approaches using the lattice equation of state describe quite well the experimental results of a multitude of hadron observables \cite{Gale:2013da}. Also the experimentally observed  fluctuation of conserved quantities has been compared in detail with the results of lattice calculation \cite{Bluhm:2014wha}, although this is quite complicated taking into account the finite size of the sytem, the finite experimental acceptance and the inelastic hadron collisions after ``freeze out", when the mean free path is too long to maintain thermal equilibrium .

Due to the sign problem lattice calculations cannot be extended to finite $\mu$. Methods, like a Taylor expansion \cite{Bazavov:2017dus} or an analytical continuation from imaginary chemical potentials \cite{Gunther:2016vcp}, have been developed to extrapolate the thermodynamical quantities away from the $\mu=0$ line but the deeper one penetrates into the finite $\mu$ region the more one enters unknown territory. Being presently outside of the range of lattice gauge calculation this is the realm of phenomenological models and subject of intensive studies. Many of these approaches suggest that the cross over, observed for $\mu=0$, continues for finite $\mu$, however with a steeper and steeper slope, before it merges finally into a critical end point followed by a first order phase transition for even larger $\mu$ \cite{Asakawa:1989bq,Stephanov:1998dy,Stephanov:1999zu}.  At zero temperature and large $\mu$ perturbative three-loop QCD calculations are available \cite{Kurkela:2016was,Vuorinen:2016pwk} which allow in this limit to compare phenomenological approaches with pQCD. 

In this paper we study the phase diagram and the thermal properties  of 3 flavour QCD at finite chemical potential in the Polyakov Nambu Jona-Lasinio (PNJL) approach, extending the work of refs. \cite{Torres-Rincon:2017zbr,Torres-Rincon:2016ahl}. This model has the merit that it is based on the symmetries of the QCD and that it is conceptually rather simple. Few vacuum observables are sufficient to fix the parameters of the theory, the extension towards finite $\mu$ is straight forward and does not need the introduction of new parameters. 
The PNJL model \cite{Fukushima:2003fw,Megias:2004hj,Ratti:2005jh,Hansen:2006ee} is  an improved version of the Nambu Jona-Lasinio model \cite{Nambu:1961fr} in which the interaction among quarks is described by a four-point interaction assuming that the gluon mass is large as compared to the momentum transfer between the quarks. Such a mass increase of gluons (and quarks) is indeed observed in nonperturbative QCD calculations. For a review we refer to  \cite{Cloet:2013jya}. 
Consequently, in NJL gluons do not appear in the quark dynamics. At low temperatures the chiral symmetry of the Lagrangian is spontaneously broken by a chiral condensate and the quarks acquire a quasi particle mass. Mesons emerge as color singlet $q\bar q$ modes. A Fierz transformation projects the Lagrangian to various quark-antiquark  and di-quark channels. The latter allows  to describe baryon masses  and to predict their temperature dependence \cite{Torres-Rincon:2015rma}. NJL type models have a long history and have been extensively used to describe the dynamics and thermodynamics of light hadrons and baryons. They offer the  possibility to study in a simple way and with a very limited number of parameters, adjusted to vacuum physics, the basic features of low temperature QCD, the basic mechanism for the spontaneous breakdown of chiral symmetry, but suffer from the absence of confinement, a consequence of the replacement of the local $SU(N_c)$ gauge invariance of QCD by a global $ SU(N_c)$ symmetry.  For more details we refer to the review articles\cite{Vogl:1991qt,Klevansky:1992qe,Buballa:2003qv}. 

In the PNJL model quarks couple to a static homogeneous Polyakov loop effective potential which is conveniently parametrized by using pure gauge lattice QCD calculation at finite temperatures  in addition to the coupling to the chiral condensate. The model reproduced successfully \cite{Ratti:2005jh} the lattice data from the Bielefeld group \cite{Allton:2003vx,Cheng:2009zi}. These calculations, however, have been further refined and nowadays all lattice groups agree on the phase diagram at zero baryon chemical potential. With these new results \cite{Borsanyi:2013bia,Bazavov:2014pvz} the standard PNJL calculations of ref. \cite{Ratti:2005jh} do not agree anymore. 

One of the advantages of the PNJL model is that it allows for a straight forward calculation of  the phase diagram in the whole $T,\mu$ plane without introducing any further parameters. To make such calculations useful it is, however, necessary to reproduce the lattice calculation at $\mu=0$. It is the purpose of this article to show that by including the next to leading order in the large $N_c$ expansion and by redefining the interaction  between the Polyakov loop potential and the quarks we obtained a good agreement with the newest lattice equation of state not only at $\mu$ = 0 but also for the expansion coefficient of the Taylor expansion toward finite $\mu$. In this article we limit ourselves to  pseudoscalar mesons. We do not expect that by including vector mesons  \cite{1105.4528,1204.3788,1207.4890,1401.4051}  the main conclusions change qualitatively.

The paper is organized as follows: we will start by describing the traditional PNJL model in section 2. In section 3, we discuss the difference of our approach as compared to the standard PNJL model. In section 4 we present our results.  First we compare our results for $\mu = 0$ with those of lattice gauge calculations, then we present the phase diagram at finite chemical potential and discuss finally the appearance of a first order phase transition for low temperatures and a finite chemical potential.  Finally, in section 5, we present our conclusions.  

	\section{PNJL Model}
		
		\subsection{PNJL Lagrangian and Polyakov loop}

The PNJL\cite{Fukushima:2003fw,Megias:2004hj,Ratti:2005jh,Hansen:2006ee,Torres-Rincon:2015rma}  model is an extension of the NJL model taking under consideration thermal gluons on the level of a mean field. The quark-quark interaction remains local, the gluons are only present as a continuous mean field surrounding the quarks.  It can be associated  to the $\frac{1}{4}F_{\mu\nu}^aF^{a\mu\nu}$ term in the QCD Lagrangian.
 We consider the Lagrangian of the PNJL model~\cite{Fukushima:2003fw,Megias:2004hj,Ratti:2005jh,Hansen:2006ee,Torres-Rincon:2015rma} 
with (color neutral) pseudoscalar and scalar interactions (neglecting the vector and axial-vector vertices for simplicity),
\bea \label{eq:lagPNJL} {\cal L}_{PNJL} &=& \sum_i \bar{\psi}_i (i \slashed{D}-m_{0i}+\mu_{i} \gamma_0) \psi_i \nn \\
&+& G \sum_{a} \sum_{ijkl} \left[ (\bar{\psi}_i \ i\gamma_5 \tau^{a}_{ij} \psi_j) \ 
(\bar{\psi}_k \ i \gamma_5 \tau^{a}_{kl} \psi_l)\right.\nn\\
&\quad& +\left.(\bar{\psi}_i \tau^{a}_{ij} \psi_j) \ 
(\bar{\psi}_k  \tau^{a}_{kl} \psi_l) \right] \nn \\
& -&    H \det_{ij} \left[ \bar{\psi}_i \ ( \unit - \gamma_5 ) \psi_j \right] - H \det_{ij} \left[ \bar{\psi}_i \ ( \unit + \gamma_5 ) \psi_j \right]  \nn \\ 
&-& {\cal U} (T;\Phi,\bar{\Phi})\ . \eea
$i,j,k,l=1,2,3$ are the flavor indices and $\tau^{a}$ ($a=1,...,8$) are the $N_f=3$ flavor generators with
the normalization 
\be \textrm{tr}_f \  (\tau^{a} \tau^{b}) = 2\delta^{ab}  \ , \ee
with $\textrm{tr}_f$ denoting the trace in flavor space.

In the Lagrangian~(\ref{eq:lagPNJL}) the bare quark masses are represented by $m_{0i}$ and their chemical potential by
$\mu_{i}$. The covariant derivative in the Polyakov gauge reads $D^\mu=\pa^\mu - i \delta^{\mu 0} A^0$, with $A^0=-iA_4$ being the temporal component of the gluon field in
Euclidean space (we denote $A^\mu = g_s A_{a}^\mu T_{a}$). The coupling constant for the scalar and pseudoscalar interaction $G$ is taken as
a free parameter (fixed e.g. by the pion mass in vacuum).

The third term of Eq.~(\ref{eq:lagPNJL}) is the so-called 't Hooft Lagrangian. It mimics the effect of the axial $U(1)$ anomaly, accounting for
the physical splitting between the $\eta$ and the $\eta'$ mesons. $H$ is a coupling constant (fixed by the value
of $m_{\eta'}-m_{\eta}$) and $\unit$ is the identity matrix in Dirac space.

 Finally, ${\cal U} (T,\Phi,\bar{\Phi})$ is the so-called Polyakov-loop effective potential used to account for
static gluonic contributions to the pressure. The Polyakov line and the Polyakov loop are, respectively, defined as
\be  L({\bf x}) = {\cal P} \exp \left( i \int_0^{1/T} d\tau A_4 (\tau,{\bf x}) \right) \ , \quad 
\Phi ({\bf x})= \frac{1}{N_c} {\rm tr}_c L({\bf x}) \ , \ee
where ${\cal P}$ is the path-integral ordering operator, and the trace ${\rm tr}_c$ is taken in the color space.

Following \cite{Ratti:2005jh} we take an homogeneous Polyakov loop field $\Phi({\bf x})=\Phi=$ const., and calculate the 
expectation values $\langle \Phi \rangle (T),\langle \bar{\Phi} \rangle (T)$ that minimize the effective potential ${\cal U} (T, \Phi,\bar{\Phi})$ at a given temperature
\bea \label{eq:minU} \left. \frac{\partial {\cal U} (T,\Phi,\bar{\Phi})}{\partial \Phi} \right|_{\Phi= \langle \Phi \rangle (T),\bar{\Phi}= \langle \bar{\Phi} \rangle (T)}
 =0 ,\nn\\ \left. \frac{\partial {\cal U} (T, \Phi,\bar{\Phi})}{\partial \bar{\Phi}}
\right|_{\Phi=\langle \Phi \rangle (T), \bar{\Phi}=\langle \bar{\Phi} \rangle (T)} =0  \ . \eea

The value of the potential for the expectation values $\langle \Phi \rangle (T),\langle \bar{\Phi} \rangle (T)$ ,  $ {\cal U} (T, \langle \Phi \rangle (T), \langle \bar{\Phi} \rangle (T))$,
gives up to a minus sign the pressure of the gluons in Yang Mills (YM) theory, corresponding to QCD for infinitely heavy quarks. The comparison with lattice gauge calculations 
for pure YM serves therefore as a guideline for the parametrization of the effective potential $U(T)$.
\be
-P(T) ={\cal U} (T, \langle \Phi \rangle (T), \langle \bar{\Phi} \rangle (T)).
\ee   
	    
	    Different possible parametrisations of the effective potential $U(T,\phi, \bar{\phi})$ have been advanced. We follow here our earlier work and use the ''polynomial parametrisation" for the YM effective potential:    

	    \begin{equation}
	    \frac{U(T,\phi, \bar{\phi})}{T^4} = -\frac{b_2(T)}{2}\bar{\phi}\phi - \frac{b_3}{6}(\bar{\phi}^3 + \phi^3) + \frac{b_4}{4}(\bar{\phi}\phi)^2,
	    \label{Potentiel polynome Polyakov}
	    \end{equation}    
	    
	    with the parameters : $b_2(T) = a_0 + a_1(\frac{T_0}{T}) + a_2(\frac{T_0}{T})^2 + a_3(\frac{T_0}{T})^3$ which are displayed in tab. \ref{tab1}. $T_0$  is the critical temperature of a Yang-Mills potential.

	    \begin{table}[h!]
	    \centerline{\begin{tabular}{|c|c|c|c|c|c|c|}
	    \hline
	    $a_0$ & $a_1$ & $a_2$ & $a_3$ & $b_3$ & $b_4$ & $T_0$\\
	    \hline
	    6.75 & -1.95 & 2.625 & -7.44 & 0.75 & 7.5 & 270 MeV\\
	    \hline
	    \end{tabular}}
	    \caption{Table of PNJL parameters for a polynomial parametrisation from lattice pure gauge fit.}
              \label{tab1}
	    \end{table}
	    
%

\subsection{Thermodynamics in the presence of the effective potential}
		All thermodynamical quantities can be obtained from the partition function here written in the path integral formalism:
		\begin{equation}
			Z[\bar{q},q] = \int\mathscr{D}_{\bar{q}}\mathscr{D}_q\left\{\int_0^\beta d\tau\int_Vd^3x\mathscr{L}_{PNJL}\right\}.
			\label{Fonction de partition}
		\end{equation}
		
		After the standard bosonisation procedure, we obtain the mean field expression of the partition function:
		
		\begin{equation}
			Z[\bar{q},q] =\exp\left\{-\int_0^\beta d\tau\int_V\frac{\sigma^2_{MF}}{4G} + Tr\ln S_{MF}^{-1}\right\}.
		\end{equation}		
		
		$\Omega(T,\mu)$, the grand potential, (we suppress here the volume dependence as we work in the infinite matter limit) is related to the partition function by:
		
		\begin{equation}
			\Omega = -T\ln(Z).
			\label{grand potentiel}
		\end{equation}
		
		We obtain for the quark part of the PNJL Lagrangian ~\cite{Hansen:2006ee,Torres-Rincon:2015rma} at the order of $(\frac{1}{N_c})^{-1}$.

		\bea
			&&\Omega_q^{(-1)} (T,\mu_i,\langle \bar{\psi}_i \psi_i \rangle,\Phi,\bar{\Phi})\nn\\ &=&\ln(Tr[\exp(-\beta\int dx^3(-\bar{\psi}(i\slashed{\partial}-m)\psi-\mu\bar{\psi}\psi))])\nn\\ &+& 2G\sum_k<\bar{\psi}_k\psi_k>^2 - 4K\prod_i<\bar{\psi}_k\psi_k> + U_{PNJL}\nn\\
 & =&  2G \sum_i \langle \bar{\psi}_i \psi_i \rangle^2
 -4H \prod_i  \langle \bar{\psi}_i \psi_i \rangle  - 2N_c \sum_i \int \frac{d^3 k}{(2\pi)^3} E_i \nn \\
&-& 2 T N_c \sum_i \int \frac{d^3 k}{(2\pi)^3} \left[ \frac{1}{N_c}\textrm{ tr}_c \log \left(1+L e^{-(E_i-\mu_i)/T} \right)
\right. \nn \\
& & \left. + \frac{1}{N_c} \textrm{ tr}_c \log \left(1+L^\dag e^{-(E_i+\mu_i)/T} \right)  \right] \ ,  
\label{OmegaMF}
\eea
where $E_i=\sqrt{k^2+m_i^2}$. Note that $G \sim {\cal O}(1/N_c)$, $H \sim {\cal O} (1/N_c^2)$ and $\langle \bar{\psi}_i \psi_i \rangle \sim {\cal O}(N_c)$. Consequently,  all terms in Eq.~(\ref{OmegaMF}) are  ${\cal O} (N_c)$. $\mu_i$ is the quark chemical potential  ($\mu_q=\mu_d=\mu_u$). In all calculations we take $\mu_s=0$.

The color traces of eq.\ref{OmegaMF} can be evaluated \cite{Torres-Rincon:2017zbr}
\bea 
 &\textrm{ tr}_c& \log \left(1+L e^{-(E_i-\mu_i)/T} \right) = \nn\\
 &\quad&\log \left[ 1 + 3(\Phi + \bar{\Phi} e^{-(E_i-\mu_i)/T}) e^{-(E_i-\mu_i)/T}\right.\nn\\
&\quad&\left.+e^{-3(E_i-\mu_i)/T} \right] \ , \\
 &\textrm{ tr}_c& \log \left(1+L^\dag e^{-(E_i+\mu_i)/T} \right) =\nn\\
 &\quad&\log \left[ 1 + 3(\bar{\Phi} + \Phi e^{-(E_i+\mu_i)/T}) e^{-(E_i+\mu_i)/T}\right.\nn\\
&\quad& \left.+e^{-3(E_i+\mu_i)/T} \right] \ . \qquad  \label{ctrace}
\eea	
To obtain the expectation values of the chiral condensate $\langle \langle \bar{\psi}_i \psi_i \rangle \rangle$ and the homogeneous Polyakov loop fields 
$\langle \Phi \rangle$ and $\langle \bar{\Phi} \rangle$, one has to minimize $\Omega$ with respect to these variables. 

		\bea
	    \frac{\partial\Omega_{PNJL}}{\partial\phi} &=& 0 \nn \\
	    \frac{\partial\Omega_{PNJL}}{\partial\bar{\phi}} &=& 0\nn \\
	    \frac{\partial\Omega_{PNJL}}{\partial\langle\bar{\psi}\psi\rangle}_q &=& 0\nn\\
	    \frac{\partial\Omega_{PNJL}}{\partial\langle\bar{\psi}\psi\rangle}_s& =& 0.
	     \label{gapeq}
	    	\eea
The indices d and s stand for up/down quarks and strange quarks, respectively.
The minimizing with respect to the condensates can be replaced by a minimizing with respect to the masses. It leads to the well known NJL gap equations ~\cite{Buballa:2003qv,Hansen:2006ee,Torres-Rincon:2017zbr}:

\be \label{eq:gap} m_i = m_{i0} - 4 G \langle  \bar{\psi}_i \psi_i \rangle + 2 H  \langle \bar{\psi}_j \psi_j \rangle  \langle \bar{\psi}_k \psi_k  \rangle \ ,
\quad j,k\neq i;\quad j\neq k \ . \ee

\begin{figure}[!h]
	\hspace*{-0.5in}\includegraphics[scale=0.28]{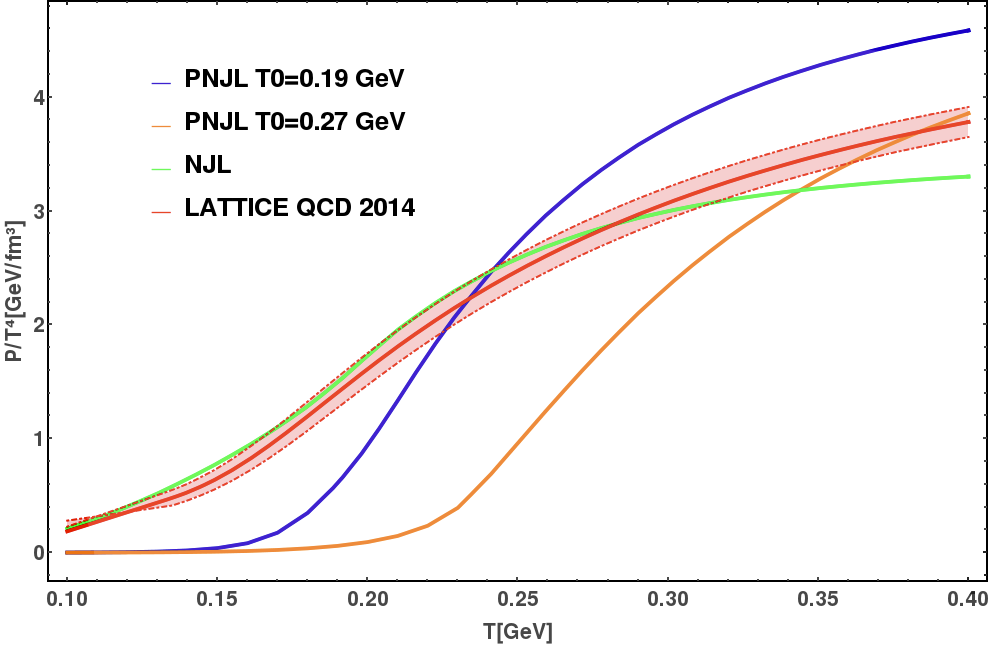}
	\caption{Comparison of the  pressure from lQCD \cite{1407.6387}, NJL and PNJL calculations. In  PNJL we use two values of $T_0$:  270 MeV (pure Yang Mills) and 190 MeV to account for the interactions between gluons and quarks \cite{Haas:2013qwp}.}
\label{eosall}
\end{figure}

The pressure is obtained by subtracting  from $-\Omega$ the vacuum pressure, taken numerically at $T=0.001 GeV$ for simplicity and $\mu = 0 \ GeV$:

\begin{equation}
	P = -\Omega(T,\mu)-\Omega(0.001,0).
\end{equation}

The pressure obtained in the NJL model, in two different parametrisations of the PNJL model and from the lattice calculations is shown in fig. \ref{eosall}. The NJL model seems to be close to the lattice results, especially at low temperature, but this is an artifact of the model. At low temperature quarks are confined and the pressure is created from hadrons.  There the Hadron Resonance Gas theory matches very well with lattice results. The lack of confinement in the NJL model leads at low T, however,  to a non vanishing pressure of the quarks.
The PNJL model has also no confinement but suppresses quark degrees of freedom at temperatures below the critical temperature.  The pressure drops quickly with decreasing  T as a consequence of the statistical confinement induced by the coupling with the Polyakov loop. At high temperature, NJL does not saturate at the same pressure as the  lattice calculations because of the lack of gluons whose contribution to the pressure is missed. The PNJL model  with $T_0=270\  MeV$, corresponding to a pure Yang Mills medium, has a vanishing pressure at low temperature because of the statistical confinement but does not match the lQCD pressure either above the critical temperature as the gluon contribution do not involve quark-gluons interaction. The PNJL calculations for $T_0 = 190\ MeV$ \cite{Haas:2013qwp}, which take into consideration the presence of quarks in the medium in an approximate way , has the same features than that for $T_0=270\  MeV$ but saturate more closely to the lattice results around the critical temperature. However, the phase transition remains too sharp to match them.

As a consequence, the reproduction of lattice results requires the addition of a mesonic pressure at low temperature vanishing naturally at high temperature and a more evolved description of the quark-gluon interaction which we will obtain using a phenomenological parametrisation of the $T_0$ parameter of the PNJL model.

\section{The grand-canonical potential in ${\cal O}(N_c=0)$ }

The next to leading order in $N_C$ of the grand-canonical potential adds in a natural way the pressure of mesons below $T_C$.
The details of the calculation of the grand-canonical potential in order ${\cal O}(N_c=0)$, $\Omega^ {(0)}_q$,  have been discussed in ref.\cite{Torres-Rincon:2017zbr}. 
Therefore we mention here only the results which are necessary for the understanding of this paper. In next to leading order the quark grand-canonical potential, $\Omega_q^{(0)}$, becomes a function of the mesons, or, more precisely, of the quark-antiquark correlation function $\Pi (\omega,{\bf p})$
\be \Omega_q^{(0)} (T,\mu_i) = \sum_{M \in J^\pi=\{0^+,0^-\}} \Omega^{(0)}_M (T,\mu_M (\mu_i))\ , \ee
where $M$ denotes the contribution of the scalar and pseudoscalar, $J^\pi=\{0^+,0^-\}$, mesons. 
$\Omega^{(0)}$ can be expressed as 
  \bea \label{eq:big} \Omega^{(0)}_{M} (T,\mu_M) &=& \frac{g_M}{2} \int \dtilde{p}  \frac{1}{2\pi i} \int_{0}^{+\infty} d\omega\nn\\ \  &\quad&\left[ 1 +\frac{1}{e^{\beta (\omega-\mu_M)}-1} +\frac{1}{e^{\beta (\omega+\mu_M)}-1} \right] \nn \\
&\times &\log \frac{  1-2 {\cal K}_M \Pi (\omega-\mu_M+i\epsilon,{\bf p}) }{1-2 {\cal K}_M \Pi (\omega-\mu_M-i\epsilon,{\bf p})} \ . \eea
and is a function of the quark-antiquark correlation function 
\bea \label{eq:twoprop} i\Pi^M_{qq'} (i\omega_m,\pvec)&=& -i T  \sum_n \int \dtilde{k}\nn\\
\textrm{Tr} \ [ \bar{\Omega}_q S_q(i\nu_n,\kvec)
 &\Omega_{q'}& S_{q'}(i\nu_n-i\omega_m,\kvec-\pvec)] \ , \eea 
where $S_q(i\nu_n,\kvec)$ is the propagator of a \renewcommand{\slash}{ \not}
quark of flavor $q$ in the Hartree approximation. The trace is to be taken in
color, flavor and spin spaces ($\textrm{Tr} = \textrm{tr}_c \textrm{ tr}_f \textrm{ tr}_\gamma$). The factor $\Omega_q$ is
\be  \Omega_q =  \unit_c \otimes \tau^q \otimes \Gamma_M \ , \ee
where $\unit_c$ is the unit matrix in color space and $\Gamma_M=\{ \unit, i\gamma_5 \}$ for scalar and pseudoscalar channels, respectively.  
The same quark-antiquark propagator $\Pi (\omega,{\bf p})$  appears in the determination of the meson mass and width.
Exploiting the Jost representation of the scattering amplitude one can obtain in the no-sea approximation (means by neglecting the first term in the brackets of eq. \ref{eq:big}) an alternative form
\bea \label{eq:Omega0next} &\Omega^{(0)}_{M} (T,\mu_M) = -\frac{g_M}{2\pi} \int \dtilde{p} \int_{0}^{+\infty} d\omega\nn\\
&\left[  \frac{1}{e^{\beta (\omega-\mu_M)}-1}
+ \frac{1}{e^{\beta (\omega+\mu_M)}-1}  \right]  \ \delta (\omega, \pvec; T,\mu_M)  \ , \eea
with
\bea \label{eq:phaseshift} \delta_M&(\omega, \pvec; T, \mu_M)  = -\frac{1}{2i} \log \frac{ 1-2 {\cal K}_M \Pi_M (\omega -\mu_M + i\epsilon, \pvec) }{[1-2 {\cal K}_M \Pi_M (\omega -\mu_M  + i\epsilon, \pvec)]^*} \nn \\
& = -\frac{1}{2i} \log |1| - \frac{1}{2}  \textrm{Arg} \frac{1-2 {\cal K}_M \Pi_M (\omega -\mu_M + i\epsilon, \pvec) }{[1-2 {\cal K}_M \Pi_M (\omega -\mu_M  + i\epsilon, \pvec)]^*} \nn \\
&= - \textrm{Arg} \left[ 1-2 {\cal K}_M \Pi_M (\omega -\mu_M  + i\epsilon, \pvec) \right]    \ . \eea
A simplifying assumption has been proposed in \cite{Hufner:1994ma,Blaschke:2013zaa} to make the argument of the phase shifts  approximately Lorentz invariant. One introduces the Mandelstam variable $s=\omega^2-p^2$, and assumes that
\bea \delta_M (\omega,{\bf p};T,\mu_M) &\simeq \delta_M (\sqrt{\omega^2-{\bf p}^2},{\bf p =0};T,\mu_M)\nn\\
&= \delta_M(\sqrt{s},{\bf p =0};T,\mu_M) \ . \eea
With this approximation  we obtain for the meson part of the grand-canonical potential
\bea \label{eq:omegafinal} \Omega^{(0)}_{M} (T,\mu_M)  = -\frac{g_M }{8\pi^3} \int dp p^2 \int \frac{ds}{\sqrt{s+p^2}} \nn\\
\left[  \frac{1}{e^{\beta (\sqrt{s+p^2}-\mu_M)}-1}
+ \frac{1}{e^{\beta (\sqrt{s+p^2}+\mu_M)}-1}  \right] \nn \\ 
\times \ \delta_M (\sqrt{s}; T,\mu_M)    \ , \eea
where the phase shift is computed using Eq.~(\ref{eq:phaseshift}).

\subsection{Interaction between quarks and gluons}

The interaction of quarks and gluons, represented by the Polyakov loop field, modifies not only the quark properties but the gluon field itself. This back reaction has been studied in ref.\cite{Haas:2013qwp} by comparing the pure Yang Mills potential $U_{YM}$ with $ U_{glue}$  obtained, when allowing for quark-antiquark excitation in the gluon propagator. The authors found that the quark-antiquark loops change $U_{YM}$ considerably. $ U_{glue}$ is related to $U_{YM}$ by
\be \label{eq:paw}\frac{{\cal U}_{glue}}{T^4} (t_{glue},\Phi, \bar{\Phi}) = \frac{{\cal U}_{YM}}{T^4} (t_{YM}(t_{glue}),\Phi, \bar{\Phi}) \ , \ee
where $t$ is the reduced temperature.  $t_{YM}$ and $t_{glue}$ are related by:
\be \label{eq:rescaling} t_{YM}=\frac{T - T_{YM}^{cr}}{T_{YM}^{cr}} = 0.57 \ \frac{T-T_{glue}^{cr}}{T_{glue}^{cr}}=0.57t_{glue} \ . \ee
 $T_{YM}^{cr}$ is the deconfinement temperature in the pure YM case (and fixed to $T_{YM}^{cr}=270$ MeV),
whereas $T_{glue}^{cr}$ is the transition temperature in the unquenched case. The numerical coefficient 0.57 is the outcome from the comparison of the two effective potentials. This procedure rescales the critical temperature from $T^{cr}_{YM}= 270$ MeV to $T_{glue}^{cr}=190$ MeV.

In this article, we go beyond a pure rescaling of the temperature due to the presence of the quarks and modify the parameters of the $U_{glue}$ by:
		
	    \begin{equation}
	    \frac{U(\phi, \bar{\phi},)}{T^4} = -\frac{b_2(T)}{2}\bar{\phi}\phi - \frac{b_3}{6}(\bar{\phi}^3 + \phi^3) + \frac{b_4}{4}(\bar{\phi}\phi)^2,
	    \label{Potentiel polynome Polyakov}
	    \end{equation}
	    
	    with the parameters : $b_2(T) = a_0 + \frac{a_1}{1+\tau} + \frac{a_2}{(1+\tau)^2} + \frac{a_3}{(1+\tau)^3}$
	    
	    where:
	    
	    \begin{equation}
	    	t_{phen} = 0.57\frac{T-T_{phen}(T)}{T_{phen}(T)}.
	    	 \label{Température réduite}
	    \end{equation}

We assume a phenomenological temperature dependence of $T_{phen}^{cr}$  of the form

	        \begin{equation}
		  T_{phen}(T) = a + bT + cT^2 + dT^3 + e\frac{1}{T},
		  \label{T0 poly}
		\end{equation}
		
and determine the coefficients a,..,e, see table II,  by comparison with lattice gauge calculations.
	    
	    \begin{table}[h!]
	    \centerline{\begin{tabular}{|c|c|c|c|c|c|c|c|c|c|c|}
	    \hline
	    $a_0$ & $a_1$ & $a_2$ & $a_3$ & $b_3$ & $b_4$ & a&b&c&d&e\\
	    \hline
	    6.75 & -1.95 & 2.625 & -7.44 & 0.75 & 7.5 & 0.082&0.36& 0.72&-1.6&-0.0002\\
	    \hline
	    \end{tabular}}
	    \caption{Full table of parameters for the PNJL model}
	      \end{table}
	      
	      \begin{figure}[h!]
	      \centering
	    \hspace*{-0.4in}\includegraphics[scale=0.6]{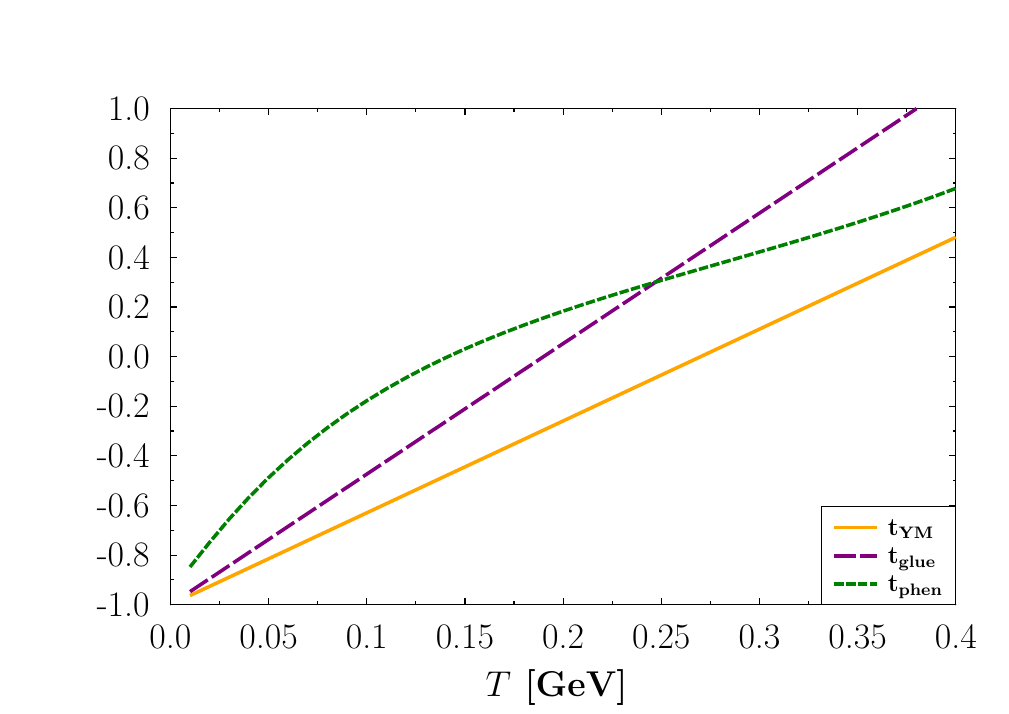}
	    \caption{$t_{YM}$, $t_{glue}$ \cite{Haas:2013qwp} and $t_{phen}$ as function of the temperature}
	   \label{efftemp}
	    \end{figure}
	      
	    This new parametrisation leaves the asymptotic limit of the gluon pressure unchanged. The rescaling is now a  function of the temperature assuming that the quark-antiquark excitation depend on the temperature of the medium. The parametrisation is taken to be polynomial.  It turns out that the fit required to go up to the third order in temperature expansion, the third and second order of expansion being especially of importance for the large temperature agreement with the lattice results. The term in $T^{-1}$ does not depend on the quark-gluon interaction. It increases the pressure at low temperature and compensates partially for the fact that  only four types of mesons ($\pi$, K, $\sigma$, $a_0$) are taken into consideration at the moment. In Fig. \ref{efftemp} we compare the different reduced temperatures for pure Yang-Mills,  from  \cite{Haas:2013qwp} Eq. \ref{eq:rescaling} and from our approach Eq. \ref{Température réduite}. We see that our reduced temperature $t_{phen}$ is higher at low temperature (where hadrons are the relevant degrees of freedom ) but comes close to the effective temperature of \cite{Haas:2013qwp} around the phase transition temperature.  It tends to the pure Yang Mills reduced temperature for high temperatures, as it should, because asymptotically we expect a plasma of noninteracting quarks and gluons.

\section{Results}

\subsection{Equation of state at vanishing chemical potential} 

As we have seen in Fig. \ref{eosall}, the PNJL approach with a constant $T_0$ and $\mu=0$ does not match the lattice equation of state.

 \begin{centering}
	    \begin{figure}[h!]
	    \centerline{\hspace*{-0.4in}\includegraphics[scale=0.6]{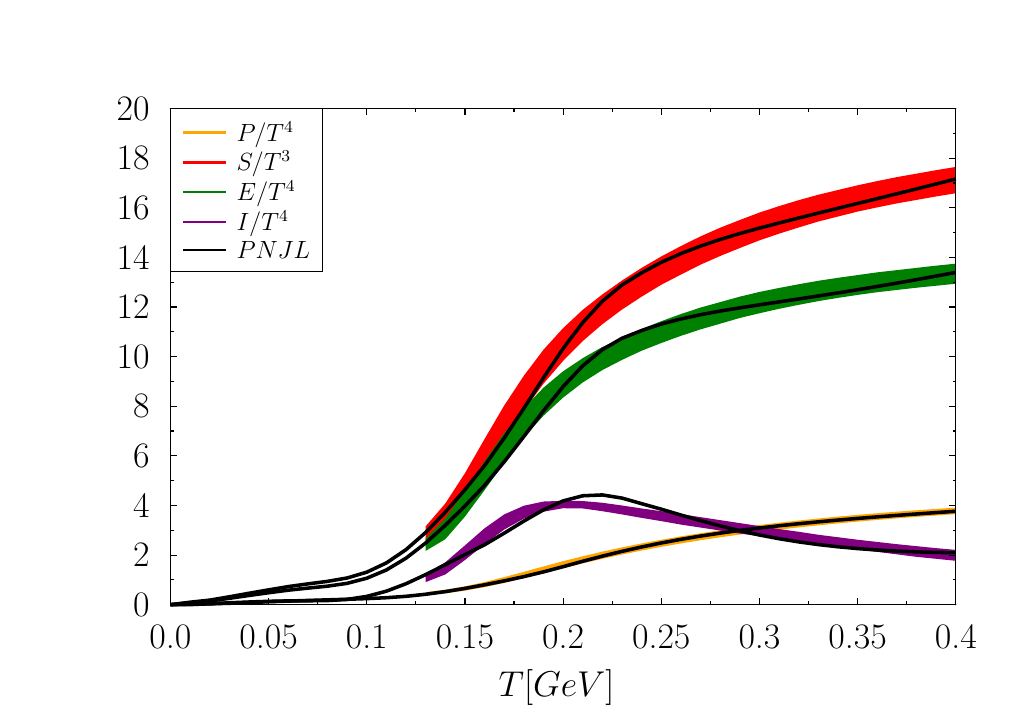}}
	    \caption{Pressure, entropy density, energy density and interaction measure calculated with  PNJL, using eq.  \ref{T0 poly},  for $\mu$ = 0. We compare our results (lines) with the lattice results (colored bands) \cite{Bazavov:2014pvz}}
	    \label{lattvsme}
	    \end{figure}
	    \end{centering}

Using the temperature dependent interaction between quarks and gluons with the parametrisation given above and taking into consideration the contribution of pseudoscalar pions and kaons and scalar mesons, $\sigma$ and $a_0$, which contribute to the  next to leading order terms of  the partition sum, we can reproduce the pressure as function of the temperature obtained  by lattice gauge calculations \cite{1407.6387}.  This is shown in Fig. \ref{lattvsme}. On the hadronic side of the phase diagram, we expect that also higher mass hadrons contribute to the pressure even if their larger mass suppresses their contribution.  Also the derivatives of the pressure as entropy, energy density and interaction measure reproduce well the results of lattice gauge calculations.
	    
 \begin{centering}
	    \begin{figure}[h!]
	    \centerline{\hspace*{0.4in}\includegraphics[scale=0.6]{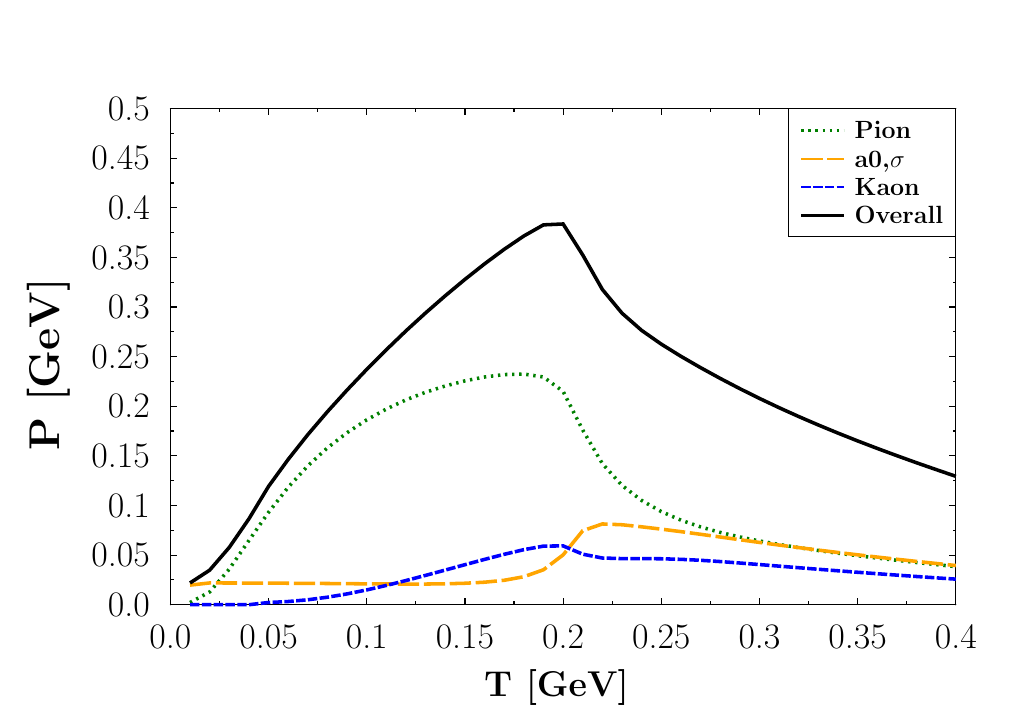}}
	    \caption{Different mesonic  contributions to the pressure at $\mu = 0$}
	    \label{Pmx}
	    \end{figure}
	    \end{centering}

Fig. \ref{Pmx} shows the contribution to the pressure of the different mesons included in our calculation. In leading order in $N_c$ neither in NJL nor in PNJL such a contribution exists.  Being the lightest meson, the pseudoscalar pion is contributing most to the pressure. Although unstable, the mesons contribute to the pressure also above the phase transition but naturally this contribution tends to vanish at large temperatures.
The pressure of the scalar mesons exhibits two different contributions. Due to its large width the scalar $\sigma$ contributes  to the pressure already at very low temperatures. At higher temperatures, we observe a chiral restoration for the scalar and pseudoscalar mesons which have there the same finite mass and consequently the same contribution to the pressure.

 \begin{centering}
	    \begin{figure}[h!]
	    \centerline{\hspace*{-0.4in}\includegraphics[scale=0.6]{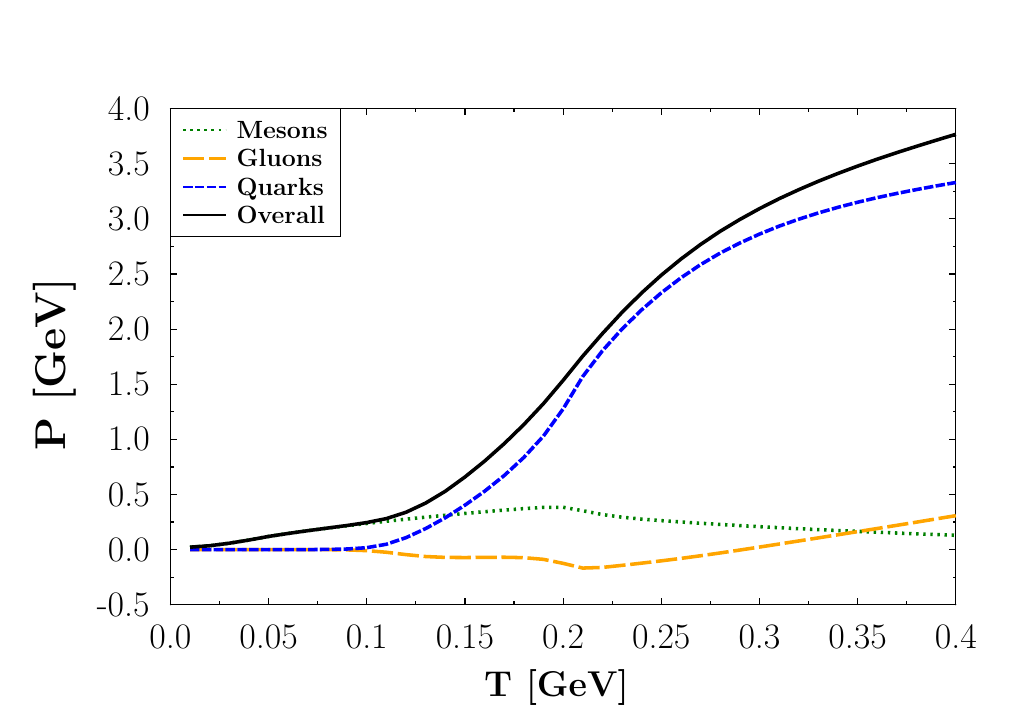}}
	    \caption{The different contributions to the total pressure at $\mu = 0$ as a function of the temperature.}
	    \label{PQ}
	    \end{figure}
	    \end{centering}

In Fig. \ref{PQ} we show the pressure contribution of gluons, quarks and mesons as a function of the temperature. At high temperatures, the pressure is dominated by the quark contribution. Mesons dominate at low temperature and present a non negligible contribution around $T_c$ indicating that at $\mu=0$ we have a cross over and not a sharp transition between the phases. The gluon pressure is negative at low temperature what can be interpreted as an attractive interaction. It becomes positive at higher temperature and reaches asymptotically the YM pressure. One notices that quarks start to contribute to the pressure below $T\approx 155\ MeV$, being usually considered as the critical temperature of the cross over phase transition. Introducing more hadrons we expect a larger hadronic contribution and therefore a smaller contribution of the quarks but one should keep in mind
that we face a cross over where the degrees of freedom in the vicinity of $T_c$ are not know even if hadron gas calculations with vacuum masses agree with  lattice gauge calculations
below $T_c$.

\begin{centering}
	    \begin{figure}[h!]
	    \centerline{\hspace*{-0.4in}\includegraphics[scale=0.6]{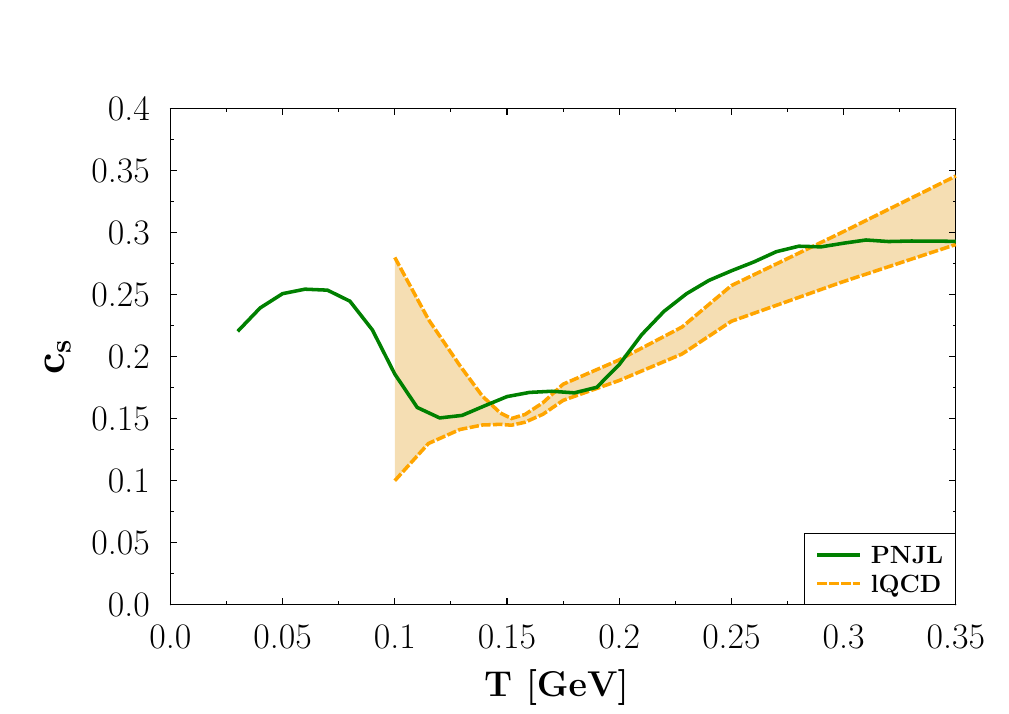}}
	    \caption{Speed of sound at $\mu = 0$.  We compare lattice calculations \cite{Borsanyi:2010cj} with our approach. }
	    \label{Pm}
	    \end{figure}
	    \end{centering}

	    Another quantity of interest is the speed of sound which is related to the compressibility of the system.
\be
c_s^2= \frac{\partial P}{\partial\epsilon}=\left(\frac{T}{S}\frac{\partial S}{\partial T} + \frac{\mu}{S}\frac{\partial N_B}{\partial T}\right)^{-1}.
\ee

Lattice calculations have found that the softest point of the equation of state, the minimum of the speed of sound is slightly below the cross over temperature \cite{Borsanyi:2010cj}. In our PNJL calculations, a second minimum appears.


The fact that not all hadrons of the hadronic spectrum are included and that the gap equations are calculated on the level of mean field implies that the parametrisation of the quark-gluon interation also compensates for this in order to reproduce the lattice pressure.
The confinement phase transition is then shifted too a lower temperature than expected to balance the chiral phase transition which is higher than the one determined with the lattice approach. This behaviour does not appear in the pressure or the entropy but only in sensible observables like the speed of sound.
We expect that the inclusion of  more hadrons and the calculation of the gap equations beyond mean field will reunified the two phase transition and localize the minimum more precisely. This would help to determine the critical temperature $T_c$ more precisely.

\subsection{Taylor expansion around $\mu=0$}

In the (P)NJL approach the extension to a finite chemical potential is straight forward. One has only to add a chemical potential in the distribution function of the quarks. We can therefore, without introducing any new parameter, calculate the thermodynamical quantities in the whole $\mu , T$ plane. To make contact with the lattice gauge calculations we can, however, also apply the same procedure by which in lattice gauge calculations the thermodynamical quantities are calculated for small but finite $\mu$.  For this we apply a Taylor expansion of the critical temperature around zero baryonic potential:
	  
	  \begin{equation}
	  	\frac{T_c(\mu_B)}{T_c(0)} = 1 - \kappa\left(\frac{\mu_B}{T_c(\mu_B)}\right)^2 + ... 
	  	\label{exp mu lqcd}
	  \end{equation}
	  
	  The $\kappa$ coefficient is \cite{1805.02960}:
	  
	  \begin{equation}
	  	\kappa = -T_c(0)\left.\frac{\partial T_c(\mu_B)}{\partial(\mu_B)^2}\right|_{\mu_B=0}
	  	\label{kappa}
	  \end{equation}

	  At $\mu_B = 0$, we get the critical temperature, determined from the inflexion point of the chiral condensate, :
	  
	  \begin{equation}
	  	T_c = 204\quad MeV.
	  \end{equation}
	  
\begin{centering}
	    \begin{figure}[h!]
	
    \centerline{\hspace*{-0.5in}\includegraphics[scale=0.4]{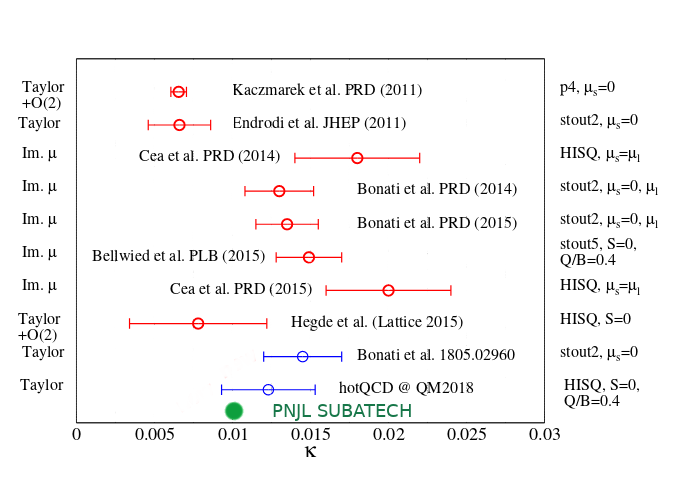}}
	    \caption{The expansion coefficient of the first order Taylor expansion for finite $\mu$ in different approaches.}
	    \label{kappatc}
	    \end{figure}
	    \end{centering}

	  The corresponding $\kappa$ coefficient is :
	  
	  \begin{equation}
	  	\kappa = 0.00989.
	  \end{equation}
	   In Fig. \ref{kappatc} this coefficient is compared with the result of lattice calculations and is found to be in good agreement with other approaches. The different approaches use partially different methods to fix this coefficient.
             Consequently, our PNJL  approach agrees with lattice data also  for finite but small chemical potentials. 
\subsection{Calculation at finite $\mu$}	 
 	    
\begin{centering}
    \begin{figure}[h!]
\centerline{\hspace*{-0.4in}\includegraphics[scale=0.6]{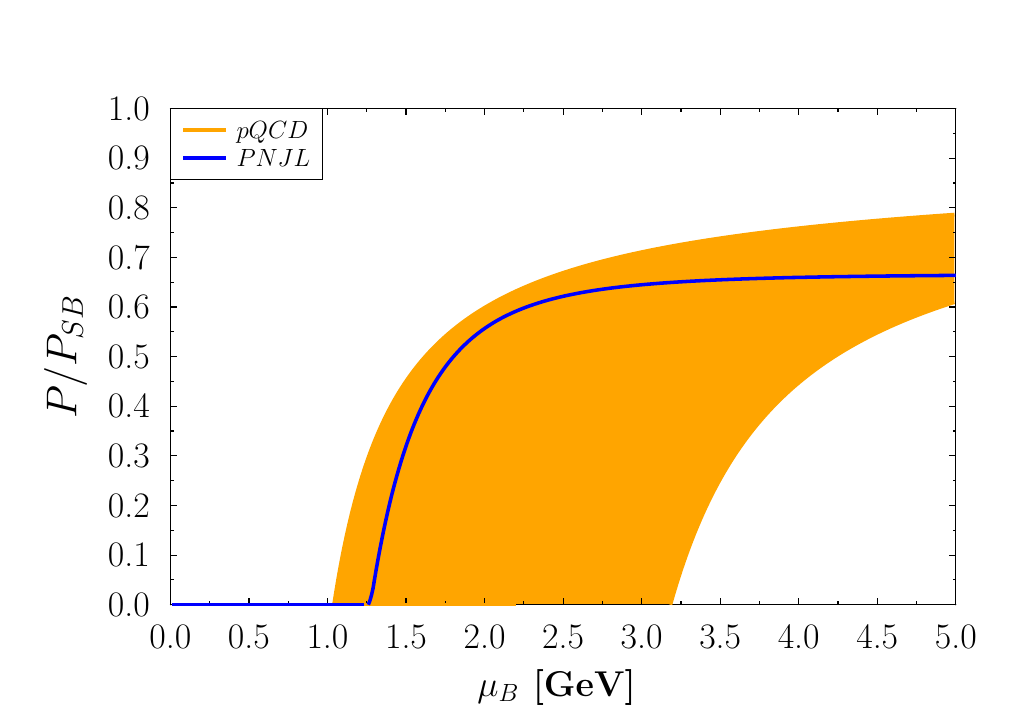}}

\caption{The quark pressure as a function of $\mu$ for a temperature of T= 0.001 GeV. We compare pQMD calculations \cite{Kurkela:2016was} (orange area) with the result of our pQMD approach (blue line).}
\label{qpres}
	    \end{figure}
\end{centering}
\begin{centering}
    \begin{figure}[h!]
\centerline{
	   \hspace*{0.3in}\includegraphics[scale=0.3]{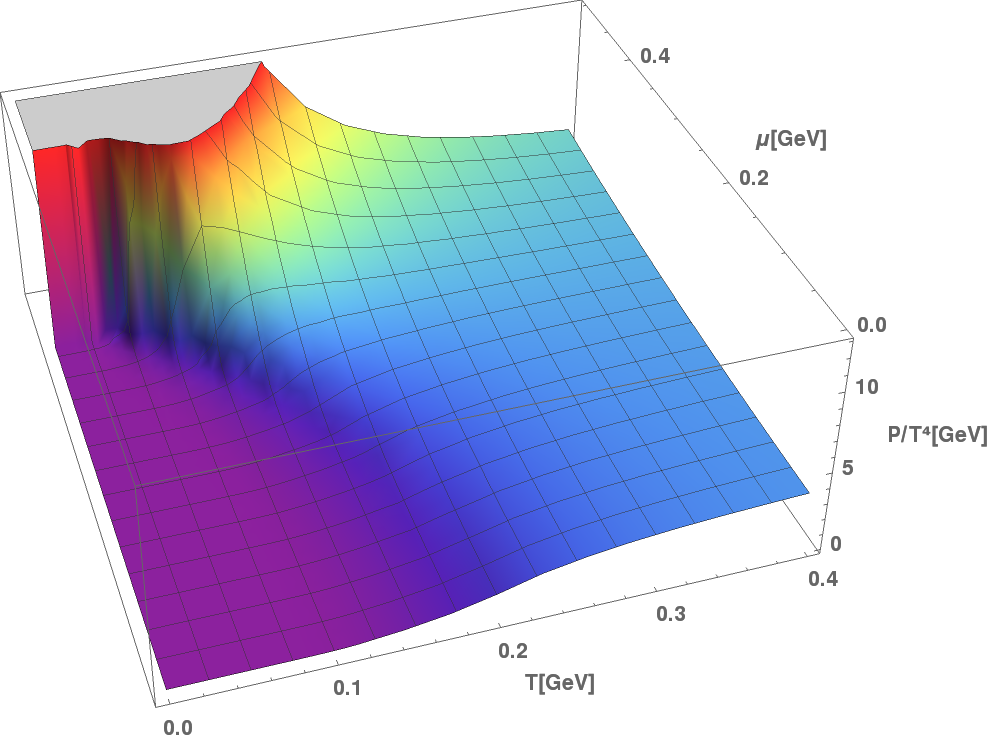}}
\caption{$P/T^4$ as a function of T and $\mu$}
\label{p3d}
\end{figure}
\end{centering}

After having verified that our approach agrees with lattice calculation for small $\mu$  we investigate the large $\mu$ limit of our approach and compare our results with perturbative QCD calculations in the hard dense loop formalism ref.\cite{Kurkela:2016was}. As seen in Fig.\ref{qpres} our approach agrees also in this limit quite well with QCD calculations.

Having verified that our PNJL approach gives the right value of the pressure for a vanishing and for large chemical potentials we have a solid basis to study  the phase diagram in between the two extremes. The result of our calculation is presented in Fig. \ref{p3d}. We see that the cross over between hadronic and quark phase continues for finite values of $\mu$, as predicted by the lattice results. With increasing chemical potential the cross over becomes steeper and steeper and finally ends up in a first order phase transition.
 The increase of  the pressure at high $\mu$ is dominated by the $\frac{1}{T^4}$ factor in $\frac{P}{T^4}$.

\subsection{Phase Transition}

\begin{figure}[!h]
		\hspace*{-0.5in}\includegraphics[scale=0.6]{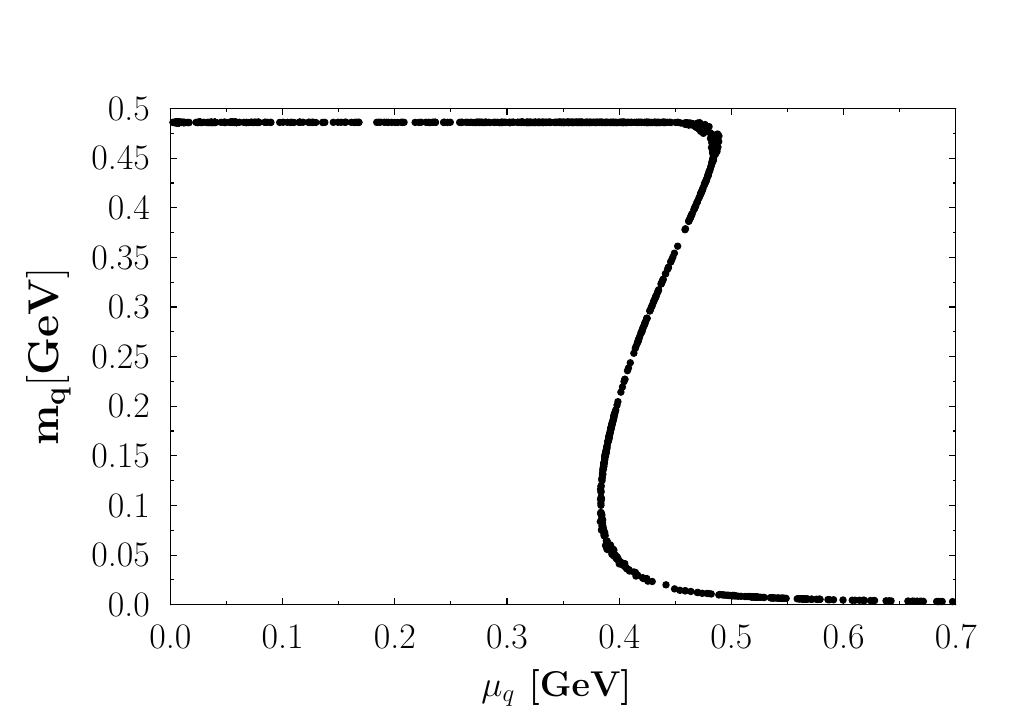}
                    \caption{Mass of the u quark around the quark chemical potential for which  a first order chiral phase transition occurs for a temperature of $ T=0.001\  GeV$.}
                     \label{mucritmas}
		\end{figure}

\begin{figure}[!h]
\hspace*{-0.5in}\includegraphics[scale=0.6]{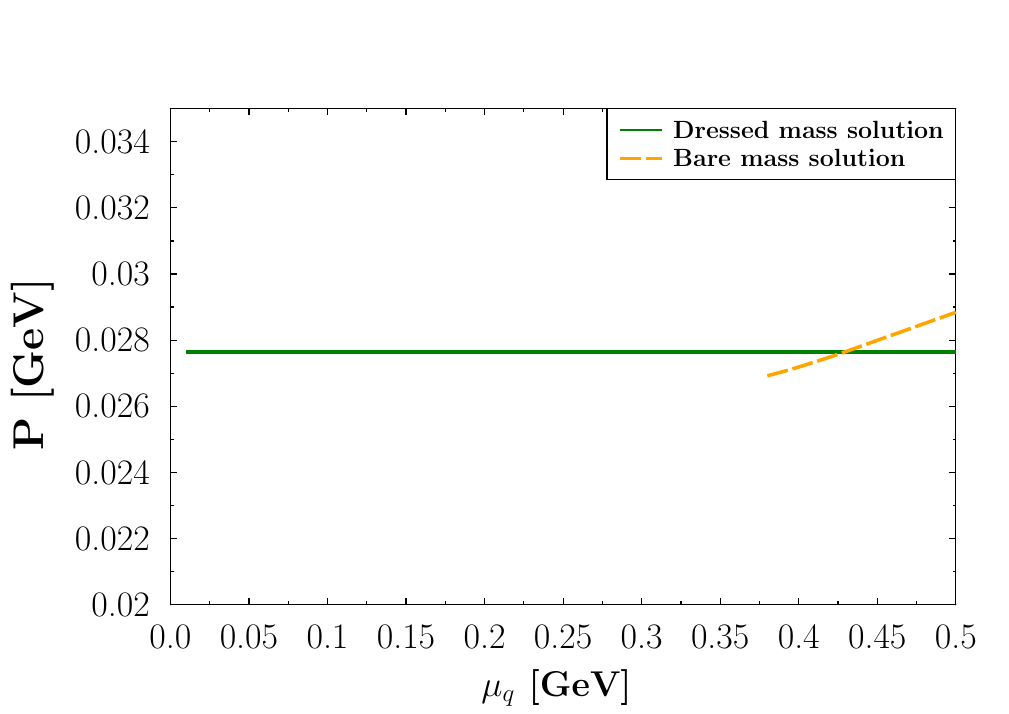}
\caption{Pressure obtain from the dressed mass (green line) sand bare mass (orange line) solutions to equation \ref{gapeq}.}
\label{mucritp}
\end{figure}

We study now the structure of this phase transition. For very low temperatures, the phase diagram is characterized by a first order phase transition at a critical quark chemical potential of $\mu_c$  =$ 0.425$ $GeV$. The order parameter of this phase transition is the quark mass. Its  dependence on the quark chemical potential is displayed in Fig. \ref{mucritmas} (for a temperature of $T= 0.001 GeV)$. We see clearly the shape of a first order phase transition. Although the critical chemical potential can be determined by a Maxwell construction it is preferable to use the Grand Potential as a function of $\mu_q$ which is displayed in Fig. \ref{mucritp}. We see there a crossing of two phases: the phase where the mass of the quarks is close to their bare mass (orange line) and in which the chiral symmetry is restored, and a phase where the quark masses are dressed because of the interactions with the medium (green line). Below the critical  chemical potential $\mu_c$  hadrons are the thermodynamically relevant degrees of freedom, above $\mu_c$  this role is played by the quarks.
 The crossing point determines the critical chemical potential $\mu_c(T)$  = $0.425$ $GeV$.

\begin{figure}[!h]
		\begin{tabular}{cc}
		\hspace*{-0.3in}\includegraphics[scale=0.6]{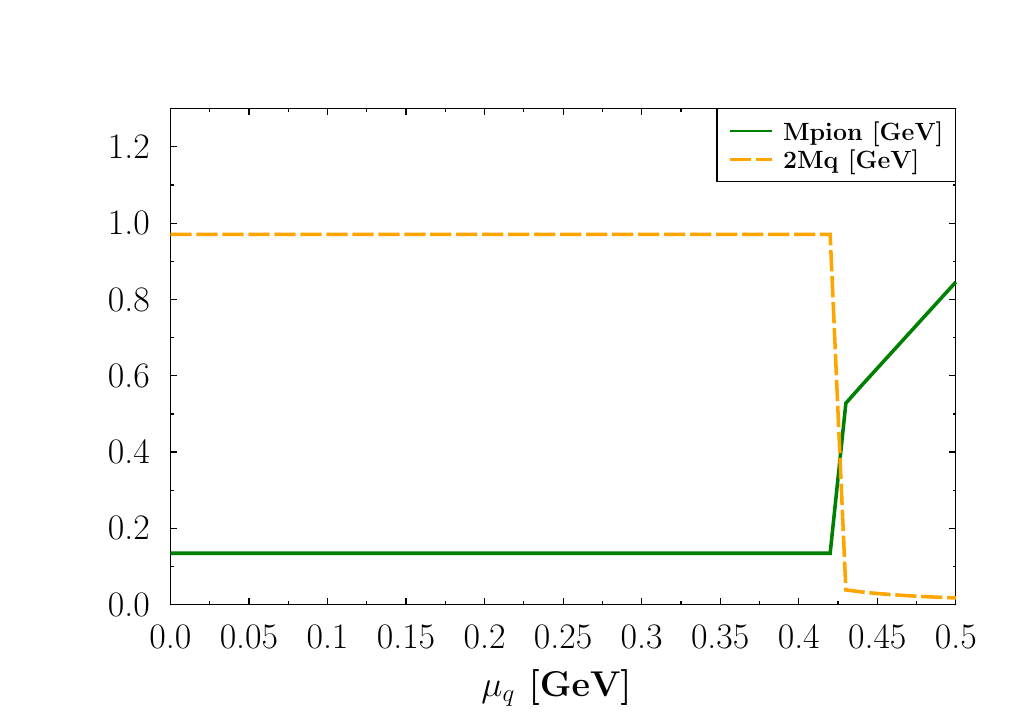}
		\end{tabular}
                    \caption{Meson and quark masses at vanishing temperature.}
                     \label{phatram}
		\end{figure}

Fig. \ref{phatram} shows the pion mass and the sum of up and down quark mass as a function of $\mu_q$. The mass of the pion, being a Goldstone boson, remains constant up to $\mu_q \approx 0.42$ $GeV$ and increases moderately for larger $\mu_q$. At the phase transition the pion mass becomes larger than the sum of the quark masses and quarks get the relevant degrees of freedom.

$\mu_c$ decreases as a function of the temperature and finally the transition becomes a cross over. The cross over region and the first order region are separated by the critical end point (CEP). The gap equations \ref{eq:gap} provides the most convenient way to calculate the critical end point. The CEP is reached when the first and second derivative of the mass with respect to the chemical potential becomes infinite.

	To simplify the system, we express everything in terms of the gap equations and we obtain a system of six equations with six unknowns: the masses of the u and s quarks, the Polyakov loop $\phi$ and its complexe conjugate $\bar{\phi}$, the temperature and the chemical potential ref.\cite{Biguet}:	
	
	\begin{align}
		g_u(\mu,T,mq,ms,\phi,\bar{\phi})&=0\nonumber\\
		 \frac{\partial\Omega_{PNJL}(\mu,T,mq,ms,\phi,\bar{\phi})}{\partial\phi} = 0 \nonumber\\
	    \frac{\partial\Omega_{PNJL}(\mu,T,mq,ms,\phi,\bar{\phi})}{\partial\bar{\phi}} = 0\nonumber\\
		g_s(\mu,T,mq,ms,\phi,\bar{\phi})&=0\nonumber\\
		\frac{\frac{\partial g_u(\mu,T,mq,ms,\phi,\bar{\phi})}{\partial mq}}{\frac{\partial g_u(\mu,T,mq,ms,\phi,\bar{\phi})}{\partial \mu}}&=0\nonumber\\
		\frac{\frac{\partial^2 g_u(\mu,T,mq,ms,\phi,\bar{\phi})}{\partial mq^2}}{\frac{\partial g_u(\mu,T,mq,ms,\phi,\bar{\phi})}{\partial \mu}}&=0\nonumber\\
	\end{align}
	
	The solution of this system is $T_{CEP}=0.11\ GeV$ and  $\mu_{CEP}=0.32\ GeV$.

\begin{centering}
	    \begin{figure}[h!]
	    \hspace*{-0.4in}\centerline{\includegraphics[scale=0.6]{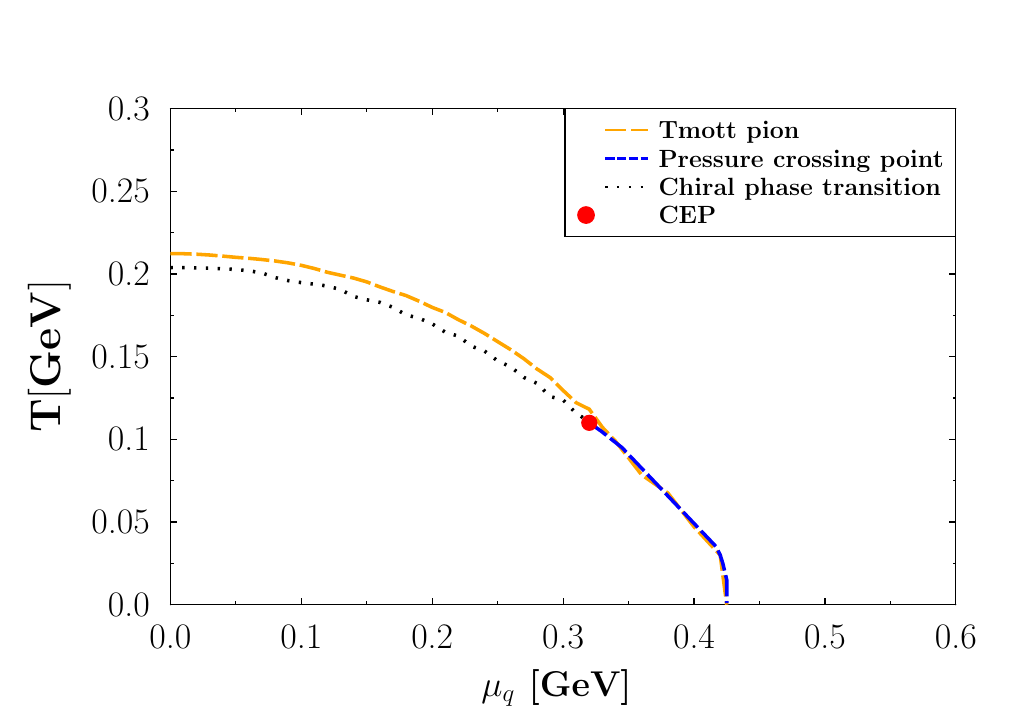}}
	    \caption{hase diagram of strongly interacting matter described by our PNJL approach.}
	    \label{Pm1}
	    \end{figure}
	    \end{centering}

Fig. \ref{Pm1} displays the relevant temperatures, as a function of the chemical potential, of our calculations. We show the Mott temperature of kaons and pions. This is the temperature at which the sum of the masses of the constituent quarks equals the mass of the meson. This temperature is a decreasing function of the temperature and very similar for kaons and pions.

Suppress this The temperature of the minimum of the speed of sound is always well below the Mott temperature and (as  the lattice calculations for $\mu=0$) below the transition temperature determined by the inflection point of the lattice as a function of the temperature.
	    \begin{figure}[h!]
	    \hspace*{-0.4in}\centerline{\includegraphics[scale=0.6]{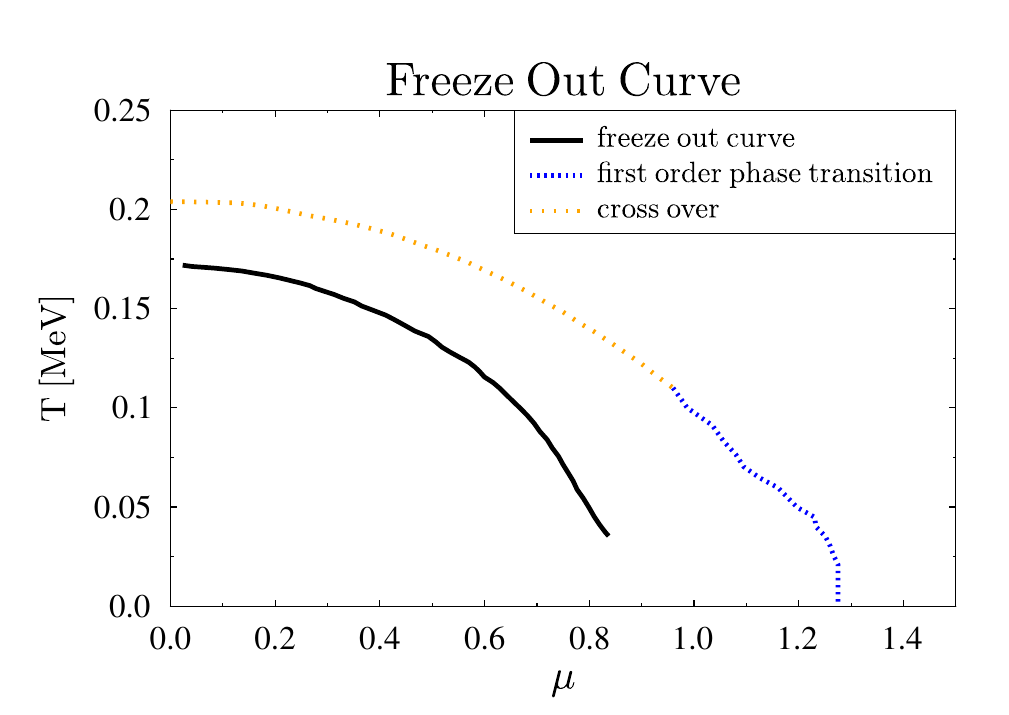}}
	    \caption{Freeze-out curve from statistical model calculations \cite{Cleymans:2005xv}. }
	    \label{Pm2}
	    \end{figure}

Can this chiral phase transition be studied by heavy ion experiments? To discuss this question we compare in Fig. \ref{Pm2} in the $T,\mu$ plane the  line of the chiral first order phase transition with the freeze out curve, calculated by Cleymans and al. \cite{Cleymans:2005xv}, which is determined by fitting the observed hadron multiplicities in the framework of a statistical model. In this approach it is assumed that after having passed the freeze-out line hadrons scatter only elastically. The point in the $T,\mu$ plane which is reached in heavy ion collisions before the system expands is not known and even whether the system comes to thermal equilibrium before the freeze-out is debated. To be consistent it has to be above the freeze-out curve. Our PNJL approach fulfills this condition and the distance between the freeze-out line and the line of the first order phase transition is small. Therefore to study this first order chiral phase transition may be in reach in heavy-ion experiments.

One word of caution should be added here. We limit us here to pseudo scalar mesons. One does not expect that vector meson \cite{1105.4528,1204.3788,1207.4890,1401.4051} change these features quantitatively but they  may change quantitatively the value of the critical temperatures. 
Our approach is also based on the quark-antiquark sector of the PNJL Lagrangian. For low temperature and high chemical potential the di-quark sector, which allows to construct baryons, will become important. To include them in the partition function will be the next step to improve the approach.
Baryons properties  have been calculated in the (P)NJL approach \cite{Torres-Rincon:2015rma} but only as a function of the temperature.
	    
 \section{Conclusion}
This work presents a improved version of the PNJL model. As compared to the standard version we have added the next to leading order (in $N_c$) contribution to the partition function.  In this order we obtain contributions from mesons and therefore the description of the thermodynamical properties below the transition temperature is largely improved. We modified also the phenomenological parametrization of the interaction between quarks and gluons which goes beyond a simple rescaling of the critical temperature $T_0$. 
Calculating the thermodynamical properties of the system we could show that for vanishing chemical potential we reproduce the results for pressure, energy density, entropy density and interaction measure  from lattice gauge calculations. Also the speed of sound agrees within the error bars with the prediction from lattice calculations even though the region of minimum shows the limit of consistency of the model regarding the fact the gap equations are calculated in mean field and consequentlynot on the same level of approximation than the pressure.

For a small but finite chemical potential our approach reproduces lattice results as well. We showed that the expansion coefficient in $\mu/T$ is in between the error bars of recent lattice calculations.

The (P)NJL calculations can be extended without any new parameter to large chemical potentials. For very large chemical potentials we compared the pressure with pQCD calculations 
and find agreement. Having verified that at small (vanishing) chemical potentials as well as at very large chemical potentials our PNJL approach agrees  with less phenomenological approaches we  can be investigate the whole T,$\mu$ phase diagram.  Our calculations predict a first order phase transition with a critical end point of
$T_{CEP}= 110\  MeV$ and $\mu_q = 320\ MeV$. Comparing our results with the chemical freeze out curve of statistical model calculations it seems to be possible that the first order phase transition can be investigated in heavy ion reactions at beam energies in the region between 3 and 10 AGeV. 
Heavy ions with these energies will become available soon at the new facilities under construction.       

This work opens as well the perspective to explore  in the finite $\mu$ region heavy ion reactions theoretically by hydrodynamical models, where the equation of state is an input, or by dynamical models, whose parameters can be calibrated to the equation of state like the dynamical quasi-particle model. Even more, the line of the first order phase transition in the $T,\mu$ plane and the freeze-out line, where, based on statistical model calculations, inelastic collisions cease, are not distant, so it may be possible to study even the first order phase transition line in heavy-ion experiments. 

This work allows also to study neutron star physics and collisions among neutron stars, both phenomena in which the equation of state plays an essential role. This will be the subject of an upcoming publication.

\section{Acknowledgment}

    We would like to thank A. Vuorinen for providing the code to compare our results with the large $\mu$ pQCD limit and for his advice as well as O. Soloveva, E. Bratkovskaya, A. Motornenko, J. Stanheimer, F. Mathieu, L. Pied, F. Cougoulic, G.Sophys, S.Delorme and M. Pierre for valuable discussions. We acknowledge funding by the Region Pays de la Loire and by the CNRS-IN2P3.

\end{document}